\documentclass[copyright]{eptcs}
\providecommand{\event}{J.~Dassow, G.~Pighizzini, B.~Truthe (Eds.): 11th International Workshop\\
on Descriptional Complexity of Formal Systems (DCFS 2009)} 
\providecommand{\volume}{3}
\providecommand{\anno}{2009}
\providecommand{\firstpage}{33} 
\providecommand{\eid}{3}
\usepackage{breakurl}        

\input{dcfs.tex}

\usepackage{graphicx,enumerate,color}

\newcommand{\noi}{\noindent}
\newcommand{\ra}{\rightarrow}

\newcommand{\Lra}{\longrightarrow}
\newcommand{\wtri}{$WT_{R_i}$}

\newcommand{\bdisplay}{\begin{description}\footnotesize\item[]}
\newcommand{\edisplay}{\end{description}}

\newcommand{\bquot}[1]{\begin{quotation}\small\noindent
  \textbf{#1}\hspace{\labelsep}\ignorespaces}
\newcommand{\equot}{\unskip\end{quotation}}

\begin{document}%
\title{The Nondeterministic Waiting Time Algorithm:\\ A Review}
\def\titlerunning{The Nondeterministic Waiting Time Algorithm: A Review}

\author{John Jack
\institute{Department of Computer Science --
  Louisiana Tech University\\ P.\,O. Box 10137 --
  Ruston -- Louisiana -- 71272 -- USA}
\email{johnjack@latech.edu}
\and
Andrei P\u aun
\institute{Department of Computer Science --
  Louisiana Tech University\\ P.\,O. Box 10137 --
  Ruston -- Louisiana -- 71272 -- USA}
\institute{Bioinformatics Department --
     National Institute of Research and Development for Biological Sciences\\
     Splaiul Independen\c tei, Nr. 296 -- Sector 6 -- Bucharest -- Romania}
\institute{Departamento de Inteligencia Artificial --
           Facultad de Inform\'atica --
           Universidad Polit\'ecnica de Madrid\\
           Campus de Montegancedo S/N --
           Boadilla del Monte -- 28660 Madrid -- Spain}
\email{apaun@latech.edu}}
\def\authorrunning{J.~Jack, A.~P\u aun}
\maketitle
%
\begin{abstract}
    We present the Nondeterministic Waiting Time algorithm.  Our technique for the simulation of biochemical reaction networks has the ability to mimic the Gillespie Algorithm for some networks and solutions to ordinary differential equations for other networks, depending on the rules of the system, the kinetic rates and numbers of molecules.  We provide a full description of the algorithm as well as specifics on its implementation.  Some results for two well-known models are reported.  We have used the algorithm to explore Fas-mediated apoptosis models in cancerous and HIV-1 infected T cells.
\end{abstract}

%
%
\section{Introduction}\label{intro}

In this paper, we will discuss a method for nondeterministic simulation of molecular signaling cascades.  The discussion centers around living systems -- e.\,g., biological cells -- and the changes in their biochemical compositions, which is brought about through the various interactions of the intracellular proteins.  Indeed, signal transduction (and molecular signaling cascades) describes the systematic interactions of different cellular proteins, starting with some sort of signal (e.\,g., an external ligand binding to a cell surface receptor) and reaching some sort of endpoint (e.\,g., the upregulation of a protein eliciting physical changes to the cell).  Our group wishes to explore the molecular mechanisms behind biochemical evolution with implied physiological responses through computational modeling.

We are at the start of a new millenium and a (relatively) new science.  We are beginning to understand ourselves on a level virtually incomprehensible at the beginning of the last century.  Humanity is on the brink of incredible technological breakthroughs -- the power to engineer/manipulate our own genomic data, the stuff that makes up who and what we are.  However, as with any engineering project, we will need computational models to be successful.  Genetic-based manipulations, changing the biochemical nature of ourselves (and our cells) for the betterment of mankind, will require an exceptional degree of certainty, which is only knowable through computer-driven modeling.  We need reliable and predictable circumstances as we go diving into our own genome.

The human genome project has given us a rich text on humanity.  The initial draft sequence of the human genome was first reported in \cite{hgp}.  So, we now have the sequence, but we lack any sort of deep understanding of it.  With our current level of knowledge, the situation is analogous to handing a tome written in Japanese to a person who understands only English and requesting that they translate the text.  There are still many uncertainties about DNA -- for instance, the precise number of genes encoded in the human genome. At first, scientists estimated that there could be two million genes in the entire human genome.  The results of \cite{hgp} put the number of genes at 30,000-40,000.  Now, the number of genes is estimated to be 20,000-25,000.  Indeed, only $2\%$ of the human genome is believed to encode for genes.


Although the number of genes encoded in a cell is alarmingly small, given the vast physical differences in the life of this planet, we look for another aspect to explain the diversity of living things.  To understand ourselves, we need to quantify the interactions of the things encoded by the genes.  The complexity of the machinery we call life stems from relatively basic components (proteins) interacting through intricate reaction networks.  Simplicity breeds complexity.

Consider the picture of a famously studied biochemical network: the EGFR network \cite{oda05}.  Looking at the picture of this network diagram, there is a disquieting moment when we realize how intensely complex and intricate even one cellular signaling pathway can be.  We are struck with a sense of awe, when we try to imagine a network diagram illustrating all of the signaling pathways at work in a normal, healthy functioning cell.  Dysfunctional cells -- for instance, a cancerous cell or one infected by a retrovirus -- do not paint pictures any less complicated.

The scientific community is attempting to unlock the molecular mechanisms underlying the functionality of the scariest and deadliest of diseases and disorders.  We do not yet understand how all of the puzzle pieces fit together.  We are only beginning to uncover and understand the components behind the most complicated -- in terms of finding a cure -- diseases and disorders.  Discovering the methods to cure diseases, such as cancer, will require an unprecedented degree of cooperation between computer scientists, mathematicians, and biologists.

In the twentieth century, we saw an explosion of breakthroughs in the physical sciences.  With the help of mathematicians and computer scientists, we were able to unlock incredible mysteries from the very large (black holes, stars, and the planets) to the very small (harnessing the power of the atom).  We have even begun venturing to other planets.  However, we are now on the cusp of a new thrust in mathematics geared towards assisting biologists.  As we turn our eyes towards the implications of genome sequencing, nanotechnology, DNA computing, and gene therapy, we see a future where predictions by computer models are a very important aspect in a fantastic new realm of science, leading to breakthroughs in therapies to fight diseases, aging, or any other ailments associated with the biochemistry of life.

Leading into the discussion of the Nondeterministic Waiting Time algorithm, we will introduce some of the existing techniques for modeling the dynamics of intracellular proteins in a reaction network.  Differential equations have been the predominate form of modeling for a very long time.  However, a promising algorithm was developed in the late 1970s using stochastic fluctuations to more accurately predict protein dynamics.  Since then, both methods have been improved, developed, modified, adapted and even combined to give us fast, accurate simulations of molecular signaling cascades.

\subsection{Modeling with differential equations}
Systems of ordinary differential equations are employed to model a wide range of phenomena, including but not limited to modeling the dynamics of molecules in a biochemical reaction network.  In fact, due to their simplicity and the speed in which they can be solved, systems of ordinary differential equations are quite popular in the modeling of molecular signaling cascades.  However, the solutions to systems of ordinary differential equations can often yield misleading results, failing to accurately represent the minority behavior of a few cells in favor of results illustrating the average behavior of the majority of cells. We will now briefly discuss how we can model biochemical systems with systems of ordinary differential equations.

We set up an ordinary differential equation for each type of molecule in the system. 
For each species~$X_i$, we have
\begin{equation}
\frac{dX_i}{dt}=f_i(X_i,...,X_n),
\label{diffeq}
\end{equation}

\noi where $f_i$'s are functions (possibly nonlinear and nonhomogeneous).  For example, the Hill function, which was initially developed to describe the binding of oxygen to hemoglobin \cite{hill10}, is a classic nonlinear function now having widespread use in describing cooperative binding (such as ligands binding to receptors).  In fact, many biological phenomena are being modeled with nonlinear functions.

Essentially, to model the dynamics of protein interactions, we need a differential equation for each protein with the functions defined according to how the proteins react to eachother in the system.  For example, we consider an early investigation into chemical equilibrium, which was made by L. Wilhelmy \cite{mcquarrie67,wilhelmy1850}.  His studies were focused on the following sucrose reaction:
\begin{equation}\label{sucrose}
H_20+C_{12}H_{22}O_{11}\ra C_6H_{12}O_6 + C_6H_{12}O_6.
\end{equation}

If we let $S(t)$ represent the concentration of sucrose, then Wilhelmy was able to show that
\begin{equation}
-\frac{dS}{dt}=kS,
\label{sucode}
\end{equation}

\noi where $k$ is a \emph{kinetic rate constant}.  He designed this equation to match the empirical evidence that the rate of decrease of sucrose concentration was proportional to the concentration remaining unconverted (the \emph{law of mass action}).


Since the time of Wilhelmy, chemical kinetics have received a great deal of attention.  To address interesting problems in the biochemistry of life, we need much more complex systems.  These systems will involve many reactions with a large degree of interdependence.  
With more complex systems, we need to approximate the solutions to the systems of ordinary differential equations.  
A popular method for approximating a system of ordinary differential equations is the fourth order Runga-Kutta method.  

We have mentioned that differential equations are not the only way to model biochemistry.  We will now briefly discuss the main stochastic technique for modeling biochemical reaction networks.

\subsection{Stochastic methods and the Gillespie algorithm}

There are situations where ordinary differential equations fail to adequately represent cellular populations.  The biochemical reason for this usually stems from situations of low molecular multiplicity.  In one of the most important works on stochastic approaches to chemical kinetics, McQuarrie \cite{mcquarrie67} provided a rich description of the historical background to stochastic techniques as well as some exactly solvable systems.

Kramers was the first to use stochastic ideas for modeling the kinetics of chemical equations \cite{kramers40,montroll58}.  The idea of stochastic approaches for modeling chemical systems revolves around the \emph{chemical master equation}.  This equation describes the probability of every possible state of the cell (with respect to biochemical composition).  Instead of a differential equation for each protein, one would essentially have a differential equation for every possible state of the cell.
For very small systems, the chemical master equation can be solved directly (see \cite{mcquarrie67} for some examples).  However, it becomes difficult or impossible to directly find the chemical master equation for systems of nontrivial size.  It is this reason which led D.\,T.~Gillespie to formulate his now ubiquitous algorithm for exactly solving the chemical master equation.

Gillespie published two landmark papers in 1976 and 1977.  In \cite{gillesp76}, he presents the framework for an exact stochastic method, which accurately predicts the chemical master equation.  Then, in \cite{gillesp77}, Gillespie describes the Stochastic Simulation Algorithm (SSA); the algorithm is now aptly named the Gillespie Algorithm, and it is the most commonly applied/adapted technique for stochastic simulation of biochemical networks.  The Gillespie Algorithm is at the heart of most discussions on stochastic modeling.

In \cite{gillesp77}, Gillespie discusses two important points on the failure of classical modeling (differential equations); the approach assumes that the time evolution of a chemically reactive system is both continuous and deterministic.  However, in nature, chemically reacting systems evolve in a discrete manner, since molecular multiplicities can obviously change only by integer amounts.  Also, it is impossible to predict the future molecular population levels through deterministic systems (a system of ordinary differential equations), because we cannot know the exact positions and velocities of all the molecules in the system.  Hence, time evolution for simulations must be a nondeterministic process, in order to account for all of the possibilities.

We will forgo an explanation of the algorithm, since it has been described in numerous publications.  However, we would like to mention the main limitation of the algorithm.  As stated in the original paper \cite{gillesp77}, the Gillespie Algorithm places a high premium on the speed of the computer's CPU.  The limitations are dependent on the number of reactions in the system.  Also, the algorithm requires multiple runs to correctly quantify the system.  This works in conjunction with the speed limitations, making stochastic simulations an enduring process.

\subsection{Improving the Gillespie algorithm}

Since its creation in 1977, the Gillespie Algorithm has been the focus for improvements in efficiency.  The most notable improvement for the Gillespie Algorithm comes from the work of Gibson and Bruck \cite{gibson00}.  The authors were able to reduce the computational complexity of the algorithm considerably, through the addition of a method for sorting the reactions and reducing the dependence on random number generation.  However, the limitation associated with reaction network growth -- number of molecules and reactions -- is still an issue.

A number of methods have now been proposed to combine differential equations with the Gillespie Algorithm.  These hybrid methods attempt to divide the reactions into \emph{fast} and \emph{slow}.  We will not provide an exhaustive discussion on each method, but we wish to mention two notable works below.

The work of Haseltine and Rawlings \cite{haseltine02} has been well-cited.  They provide the theoretical background for dividing reactions into fast and slow subsets, allowing for the fast reactions to be approximated either deterministically or as Langevin equations.  Essentially, they are able to integrate the system over much larger time steps than the original Gillespie Algorithm.  For the original Gillespie Algorithm, increasing the number of molecules for a fast-reacting protein will significantly increase the computational load; however, by using deterministic processes for fast reactions, the computational load of their algorithm will not increase in this case.

Rao and Arkin \cite{rao03} applied the quasi-steady state assumption to modify the Gillespie Algorithm.  Using the quasi-steady state assumption, they were able to reduce model complexity by reducing the number of molecular species and reactions.  Essentially, the assumption is that the net rate of formation is approximately zero for highly reactive and transitory species -- e.\,g., enzyme-substrate complexes.  In their paper, the authors provide some mathematical rigor behind the algorithm as well as some results for example systems.

\subsection{Our work}
In this paper, we provide the framework for the Nondeterministic Waiting Time algorithm.  The algorithm is designed to run faster than a Gillespie-based approach, yet maintain a level of nondeterminism separating it from approaches based on ordinary differential equations for systems of low molecular multiplicity.  The nondeterminism stems from reactions competing over limited reactants.

In Section \ref{algorithm}, we provide the motivation for the algorithm, followed by a full description of the algorithm and a discussion on its implementation.  In Section \ref{results}, we show some results for two example systems, illustrating the difference between our algorithm and ODE-based or Gillespie-based modeling.  Finally, Section \ref{conclusions} contains our conclusions and future research interests.




\section{The nondeterministic waiting time algorithm}\label{algorithm}

In Section \ref{intro}, we provide a survey -- but in no way an exhaustive account -- of the numerous algorithms pertaining to the simulation of molecular interactions in a biochemically reactive system.  Each of the simulation techniques discussed have their own particular strengths and weaknesses.  For the stochastic techniques, the strengths and weaknesses typically revolve around the accuracy with which the algorithm can predict the chemical master equation and the computational efficiency with which it functions.  For the deterministic techniques, speed is one of the main strengths, but at a loss of the randomness inherent in living biochemical systems.  Predicting the behavior of the minority populations is impossible with deterministic techniques, since they display the average behavior of the system and, thus, favor the majority behavior.

In this Section, we will discuss a different type of biochemical simulation algorithm.  A technique designed in such a way that it is capable of exhibiting behavior similar to continuous deterministic approaches for certain biochemical models, but it behaves similar to the discrete stochastic approaches -- e.\,g., the Gillespie Algorithm -- for certain other types of systems.  We have chosen to call this technique the Nondeterministic Waiting Time (NWT) algorithm.

The NWT algorithm is an extenstion of the previous modeling efforts from the lab
of Andrei P\u aun.  In~\cite{cheruku07} the groundwork for a Deterministic Waiting Time (DWT) algorithm was laid out.  Here, we provide an efficient, refined algorithm with several important extensions from the previous algorithm.  Notably, the algorithm now has a nondeterministic component and a memory enhancement, which gives it a unique ability to simulate reaction competition over limited numbers of reactants.  Additionally, the implementation of a min-heap with special maintenance functions improved the efficiency of the previous simulation algorithm.

\subsection{Introducing membrane systems}\label{intromemsys}

We will present the foundations for our discrete, nondeterministic biochemical simulation techinque: the NWT algorithm.  In the design of this algorithm, our goal is to define a simulation technique between the realm of the Gillespie Algorithm and modeling with systems of ordinary differential equations.  Moreover, we wish to have a modeling technique which is less computationally intensive than the Gillespie Algorithm, yet maintains a level of nondeterminism (or stochasticity) which sets it apart from solutions to systems of ordinary differential equations.  To describe the foundations of the NWT algorithm, we must explore the realm of a relatively new paradigm of computing: Membrane Systems (or P Systems).

The evolution of DNA, RNA, and proteins during life's tenure on Earth is the story of the storage and application of information, similar in development to the field of computer science.  For these macromolecules of life, there is a classic debate between life scientists surrounding which of them evolved first.  Implicit to this debate is an emphasis on the billions of years of information theory inherent in life and nature.

Let us assume, for the sake of argument, that RNA existed before the others.  Then we can imagine that the initial molecules of life had the ability to store information, the way modern messenger RNA (mRNA) carries the genetic information from the nucleus to the cytoplasm for translation into a protein.  However, we can also imagine some initial molecules of life possessing the ability to put the information to work, the same way small nuclear RNA (snRNA) has responsibilities in the modern cells pertaining to the transcription of a gene.  It is from the incredible achievements in information storage and application, apparant in all living cells, that Membrane Systems evolved as a way to view the molecular activity within a cell as a computation.

The concept of computing with membranes was first proposed in 1998 by Gheorghe P\u aun \cite{gpaun98}.  As a model for computation, Membrane Systems have proven to be quite useful; they take advantage of exponential space in order to solve computationally hard problems efficiently.  In their short history as a computational paradigm, a multitude of Membrane Systems have been proposed.   These Membrane Systems have the ability to attack NP-Complete problems through the use of exponential space, sharing some of the fundamental concepts of biological parallelism with DNA computing \cite{adleman94}.  For example, a Membrane System has been described which is capable of solving the boolean SAT problem in linear time \cite{ferretti00}. However, no one has yet been able to build an actual Membrane Systems computer out of a cell.

If we could harness the computational power of a cell, we could break through the glass ceiling on efficient solutions to computationally complex problems -- i.\,e., NP-complete.  Although the majority of Membrane Systems research has been on abstract models and theory, there are a few groups who wish to use Membrane Systems in a different way.  Some computer science groups are investigating the use of Membrane Systems to address problems in computational biology.  This is the direction of our interest.

For our purposes, we will define a Membrane System, $\Pi$, in the following way:
\begin{equation}
\Pi = ( \Sigma, L, \mu, M_1,..., M_m, R_1,..., R_m),
\label{Membranesys}
\end{equation}

\noi where
\begin{itemize}
\item The alphabet, $\Sigma$, is a nonempty set (of proteins).
\item A set of labels, $L$, representing all of the different compartments of the system.
\item The membrane structure, $\mu$, represents the hierarchical organization of the 
different compartments,~$L$.
\item The multiplicity sets, $M_i$ where $1\leq i\leq m$, contain the multiplicities of the proteins within each compartment -- i.\,e., the number of molecules per protein.
\item The rule sets, $R_i$ where $1\leq i\leq m$, contain the rules associated within each compartment.  The rules are the chemical reactions.
\end{itemize}

It is worth mentioning that for the remainder of this text, reaction/rule and species/protein will be used interchangeable.  One of the nice features of Membrane Systems is that they can easily be comprehended using a graphical representation.




The rules of the system govern the biochemical evolution of the system.  The rules act in a maximally parallel manner -- i.\,e., for the system to evolve from one state to the next state, all rules which can be applied are applied.  For abstract Membrane Systems, the transition from one system state to the next is called a computation.  The computations of a Membrane System are similar to the transitions of a Turing Machine.  The system continues until it reaches a stopping configuration.  



For our NWT algorithm, we would like to use Membrane Systems as the framework.  The simulation of a chemically reactive system -- e.\,g., a living cell~-- is merely the evolution of the Membrane System according to the set of rules, $\lbrace R_1,...,R_m\rbrace$.  To do so, we will need a discussion on how/when rules will execute.   As the rules describe the interactions between proteins in the alphabet, the evolution of the Membrane System tracks protein dynamics.  While there are many types of biochemical reactions, we will list a few of these to facilitate our understanding on the design and implementation of the Membrane System proposed in Equation \ref{Membranesys}.  Some basic examples of biochemical reactions are listed in Table \ref{basicex}.

{\renewcommand{\arraystretch}{1.2}%
\begin{table}[h!b!p!]
\centering
\caption[Typical examples of biochemical reactions]{Typical examples of biochemical reactions.} \label{basicex}
\begin{tabular}{lc}
\hline
{\bf $\mathbf{R_1}$: Monomolecular decay:} & $A\overset{k_d}\Lra \emptyset$\\
{\bf $\mathbf{R_2}$: Monomolecular reaction:} & $A\overset{k_x}\Lra B$\\
{\bf $\mathbf{R_3}$: Bimolecular reaction:} & $A+B\overset{k_y}\Lra C$\\
{\bf $\mathbf{R_4}$: Trimolecular reaction:} & $A+B+C\overset{k_z}\Lra D$\\
\hline
\end{tabular}
\end{table}}

In order for the Membrane System to illustrate protein dynamics over time, we need to discuss the temporal aspects of the rules of the system.  To model the biochemistry of life, the individual chemical reactions described in the system occur must over different lengths of time in an asynchronous manner.  The rules of our Membrane System obey the \emph{law of mass action}, which was first formalized in 1864 \cite{guldberg1864ii,guldberg1864i,waage1864}.  The law of mass action states that a reaction rate is directly proportional to the number of reactants available in the system.  In other words, the time a reaction takes to occur is dependent on the number of its reactant molecules.

With the law of mass action, we have a way to associate time dynamics with the evolution of the Membrane System as it jumps from one configuration to the next.  We also have a way to make the rules occur in an asynchronous manner.  As previously mentioned, a Membrane System typically evolves by applying rules in a maximally parallel manner.  In other words, when the system jumps from one configuration to the next, any and all rules which can be applied (given sufficient reactants) are applied.  But, when we use the law of mass action, the reactant concentrations govern how much time must transpire before a particular reaction can take place.  Therefore, for any particular configuration of the Membrane System, the number of reactant molecules for a given reaction determines when that reaction is next slotted to occur.  The values associated associated with the law of mass action are called \emph{kinetic rates}.

These kinetics rates must be determined through biological experimentation.  As such, the kinetics of a chemically reactive system are often described as concentration-based values.  The reason for this is the fact that biological results are often generated from enormous populations of cells.  Often a biological experiment will consider a population of millions of cells.  To determine the intracellular concentrations of proteins, these cells are then lysed as a large population.  The intracellular molecules are then measured in terms of light intensity (radiological or photonic markers), which gives data on general concentrations of particular molecules across the population.  Finally, the values are averaged to give the concentration per cell.  Therein, lies a major problem with biochemical modeling.  We rely on the values generated in the biological lab, and these values are often generated over entire cell populations instead of individual cells.  Hence, the interesting phenotypic, biochemical and physiological characteristics of individual cells can be lost in lieu of the behavior of the majority of the cells in the population.

There are techniques which can measure single cell dynamics, and this technology is very promising.  For instance, important results on p53 have been reported~\cite{lahav04} from the results of measuring single cell dynamics instead of averaging over cellular populations.  The authors of \cite{lahav04} were able to show individual cells undergo not dampened oscillations, as previously reported \cite{baror00}, but each individual cell instead undergoes different numbers of oscillations.  The average behavior for the cell population appears dampened, but individual cells do not.

At Louisiana Tech University, we are collaborating with Mark DeCoster's biomedical laboratory in order to study single cell data via a high-speed imaging system.  It is our hope that future collaborations between Andrei P\u aun's computational group and Mark DeCoster's biomedical laboratory will help unlock some of the secrets behind Fas-induced apoptosis.  Regardless of whether data comes from large cell populations or single cell dynamics, we, as modelers, must remain vigilent and build the best models with the data available to us.

The kinetic rates, $k_R$ for some reaction $R$, will often have units based on nMs, $\mu $Ms, etc.  Let's assume we have these kinetic rates for every reaction in our Membrane System.  We have already stated that our technique is a discrete one.  Therefore, in order deal with multiplicities of proteins as opposed to concentrations, we must calculate a discrete kinetic constant.  This discrete kinetic constant will be based on numbers of molecules.  When we initialize the Membrane System, we must calculate the discrete kinetic constants from the concentration-based kinetic rates in the following way
\begin{equation}
\mathit{const}_R = \frac{k_R}{V^{i-1}\times N_A^{i-1}},
\label{constr}
\end{equation}

\noi where $V$ is the volume of the system, $N_A$ is Avogadro's constant ($6.0221415\times 10^{23}$) and $i$ is the number of reactants involved in the reaction.

With the law of mass action and the discrete kinetic constants, we have the means to allow the rules of the Membrane System to occur at times dependent on reactant multiplicities, which are subject to variation throughout the entire simulation run.  We can now define a reaction's \emph{Waiting Time} (\emph{WT}).  The Waiting Time is a value associated to each reaction, signifying when the next time a single instance of the reaction will occur.  As molecular multiplicities will change throughout a simulation, so will the reaction Waiting Times (in accordance with the law of mass action).


For a first order reaction, like $R_1$ from Table \ref{basicex}, the Waiting Time is calculated with the following equation:
\begin{equation}
WT_{R_1}=\frac{1}{k_d*|A|},
\label{wt1}
\end{equation}

\noi where $A$ is the reactant required for reaction $R_1$, $|A|$ represents the number of molecules present in the system at the moment of Waiting Time calculation, and~$k_d$ is the discrete kinetic constant.  N.\,B., $R_1$ and $R_2$ from Table \ref{basicex} are calculated the same way (replacing $k_d$ with $k_x$) because they both have reaction order one and use the same reactant species, even though the products are different.

If one of the reactants for a reaction has no molecules present in the system, then we set the Waiting Time equal to infinity; since we have chosen to implement the algorithm in ANSI C, this is can be easily accomplished as $\frac{1.0}{0.0}=\infty$.  For higher order reactions, we need to incorporate the other reactants into the calculation of Waiting Time.  Following the examples in Table \ref{basicex}, a second order reaction (bimolecular) would be calculated in the following way
\begin{equation}
WT_{R_3}=\frac{1}{k_y*|A|*|B|},
\label{wt2}
\end{equation}

\noi and a third order reaction (trimolecular) would be
\begin{equation}
WT_{R_4}=\frac{1}{k_z*|A|*|B|*|C|},
\label{wt3}
\end{equation}

\noi where $A$, $B$, and $C$ are the reactants required for reactions $R_3$ and $R_4$, $|A|$, $|B|$, and $|C|$ represent the number of molecules present in the system at the moment of Waiting Time calculation, and $k_y$ and $k_z$ are the discrete kinetic constants.

In this way, the initial Waiting Time is calculated for every reaction in the entire system.  Now the question remains: how do we efficiently sort the reactions so that we can easily determine which reaction is slotted to occur next?  To do this, we will need to build a min-heap (based on reaction Waiting Times), where the top of the heap is the reaction with the smallest Waiting Time -- i.\,e., the next reaction to occur.   However, we will not be able to maintain the min-heap, as the Waiting Times change, in a standard manner.   When a rule is applied, multiple nodes can have changes to their Waiting Time, since the multiplicities of the system are changed.  Thus, multiple Waiting Times can fail the min-heap property throughout the tree simultaneously after each time step.  In order to efficiently evolve the Membrane System, we will need to incorporate some special heap maintenance functions, similar to those proposed by Gibson and Bruck \cite{gibson00} in their modification of the Gillespie Algorithm.


\subsection{Description of the algorithm}\label{NWT}

We have a Membrane System, which describes all aspects of the system -- e.\,g., rules, compartments, protein types, numbers of molecules per protein, etc.  We discussed the fact that our Membrane System will not evolve in a typical (maximally parallel) manner, because the reactions occur in an asynchronous manner over discrete time intervals of different lengths according to the law of mass action.

Next, we provide a description of the NWT algorithm.  The Membrane System evolves through the execution of reactions in a Waiting Time-dependent manner until a desired simulation time has been reached.  We will now list the Steps for the NWT algorithm.

\begin{enumerate}[{\sc(i)}]
\item  {\it Build Membrane System:} Import data for Membrane System -- alphabet, membrane 
hierarchy, etc.  Convert protein concentrations to molecular multiplicities.  Convert 
kinetic rates to discrete kinetic constants.  For each reaction $R_i$, where 
$1\leq i\leq m$, we calculate the initial Waiting Time, \wtri  .  
Choose the desired amount of time for the simulation, $\tau_{fin}$.  Set current simulation
time to zero ($\tau = 0$).
\item {\it Build Heap:}  Using the reaction Waiting Times, we build a min-heap of all 
reactions in the system.
\item  {\it Select Rule:} Choose the reaction with the lowest Waiting Time -- the top of the 
min-heap.  Upon selecting the top node, recursively check to see if there are any children 
nodes sharing the minimum Waiting Time.  If such a tie for minimum Waiting Time exists, 
proceed to Step IV.  If no tie exists, then proceed to Step V.
\item  {\it Handle Tie:} Check the multiplicities of the reactant species for all tied 
reactions.  If there are enough reactants to satisfy all of the reactions with the minimum 
Waiting Time, implement all tied reactions.  If there are not enough reactants to accommodate 
all the reactions, use the nondeterministic logic to apply as many rules as possible.
\item  {\it Apply Rule:} Update the multiplicities of the reactant(s) and product(s) for the 
reaction(s) from Step III.
Aggregate the simulation time (\hbox{$\tau = \tau + WT_{applied}$}).
\item  {\it Update Rules:} Recalculate the Waiting Time for all reactions whose reactants 
include the products or reactants of the applied reaction(s).  That is, we need to see how 
the multiplicity changes from the applied reaction(s) have affected the Waiting Times for 
all rules dependent on those proteins with changed multiplicity.  For each such reaction 
compare the new Waiting Time with the existing Waiting Time and keep the smallest of the
two (unless the new time is infinity).
\item {\it Memory Enhancement:} If the recalculation of a reaction's Waiting Time results 
in a value of infinity, then we must store the amount of time waited as a percentage 
($\mathit{Mem}_{\mathit{perc}}$).  If the recalculation of a reaction's Waiting Time 
results in a real value and the previous value was infinite, then the Waiting Time will
need to be adjusted according to the stored memory percentage.
\item  {\it Heap Maintenance:} Adjust the min-heap, bubbling reaction nodes up or down in 
order to satisfy the min-heap property, once reaction Waiting Times have been recalculated
according to the multiplicity changes.  N.\,B., to accomodate the multiple changes in 
Waiting Times, we employ nonstandard heap maintenance methods.
\item  {\it Termination:} If $\tau = \tau_{fin}$, then terminate the simulation.  
Output the multiplity information for entire simulation.  Otherwise, go back to Step~III.
\end{enumerate}

To initialize our simulator, we use a file encoded in the Systems Biology Markup Language (SBML).  SBML is one of the most popular methods to encode biochemical models, developed through a broad international collaborative effort involving the cooperation of many institutions \cite{hucka03}.  We chose SBML for its visibility and availability.  SBML has in place an extensive emailing group for quick discussions on coding issues and future extensions/developments of the standard.  

To generate the SBML files, we use the CellDesigner software \cite{funahashi03,funahashi08} which is also the result of a large international collaborative effort but maintained through Keio University.  The CellDesigner software provides an easy graphical interface with which to program the models.  CellDesigner has many functions, including simulation packages, but we are only concerned with its ability to generate SBML models (pictures and code) through a simple, user-friendly graphical interface.

Using the SBML code, we can populate every aspect of the Membrane System.  We have chosen to implement the algorithm in the C programming language.  Initially, we programmed the entire algorithm in Java for portability reasons. However, for larger models, like those we describe in \cite{jack07,jack08,jack08hiv}, we found the speed benefits of C to be necessary in reducing simulation runtime.  Also, C gave us the ability to parallelize our simulations via MPI.  To continue our discussion, we will be using the word struct.  However, a Java implementation could be understood by substituting the word struct with object.

\begin{figure}[ht]
\centering
\includegraphics[scale=0.4]{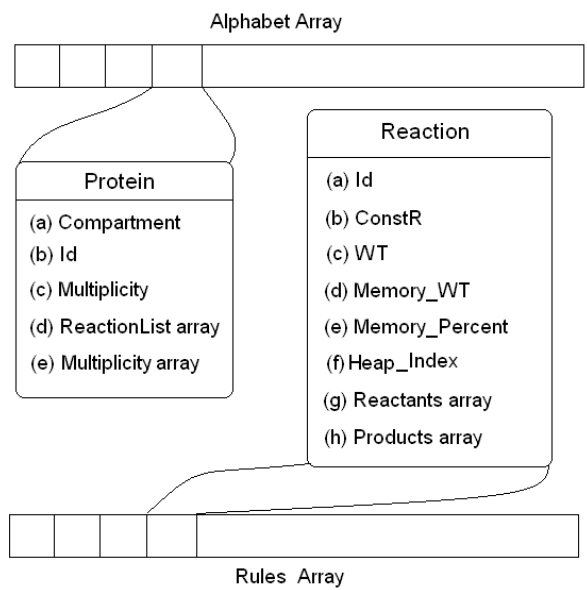}
\caption{The two structs: Protein and Reaction nodes used to build the Alphabet and Rule sets of the Membrane System.}
\label{thestructs}
\end{figure}

All of the elements of the Membrane System can be contained in two arrays of structs.  We refer to one array as the ``Alphabet'' and the other array as ``Reactions''.  We represent these structs graphically below (Figure \ref{thestructs}).  By explicitly explaining these two arrays and the aspects of the structs, we can better understand their relationships and how we have effectively and efficiently implemented the NWT algorithm.

For the Protein struct, the first three components consist of (a) a string, (b) a string, and (c) an integer (the number of molecules must be a whole number).  The Compartment and Id components are fixed throughout the entire simulation run.  The Multiplicity component is subject to change throughout the simulation, increasing or decreasing by one whenever a given protein is the product or reactant of an applied reaction.

Part (d) of the Protein node is the ReactionList array.  The ReactionList array contains the indices of all of the reactions for which the protein is a reactant.  The purpose of the ReactionList array is O(1) time access to all reactions which use the protein as a reactant.  This is convenient for quick recalculation of the Waiting Time of a reaction, after a change in the multiplicity of at least one of its reactants has occured as the result of the execution of a rule -- that is, Step VI of the NWT algorithm.

Finally, component (e) is the Multiplicity array.  At each second (easily modified to handle minutes or hours as needed) throughout the entire simulation run, the multiplicities of all proteins are saved to their Multiplicity array.  This allows us to circumvent read/write access of the slower memory (IDE/SATA harddrives) in favor of faster memory (RAM).  
The Multiplicity array is dumped to a results file every time the array becomes full, then it begins filling again, continuing until the desired simulation time is reached.

Next, we will consider the components of the Reaction structs. Component (a) comes directly from the SBML code and is stored as a string.  We discussed earlier (Equations \ref{constr}, \ref{wt1}, \ref{wt2}, and \ref{wt3}), the calculation of the (b) $ConstR$ and the (c) $WT$.  Both of these values need to be stored with double precision.  This is due to the fact that most biochemical reactions take less than one second to occur.  Hence, we want to use the largest standard type definition to ensure the largest possible number of decimal places.  Values (a) and (b) do not change after initialization.  However, as discussed, the Waiting Times are recalculated as the multiplicities of the reactants change.

Components (d) and (e) deal with the memory enhancement of the NWT algorithm.  These parts are better left unexplained until the discussion of memory enhancement in Section \ref{memenhance}.  We would like to note that these values are stored with double precision.  The HeapIndex (f) is required to handle heap maintenance.

Finally, the Reactants array and the Products array (components (g) and (h)) contain the indices of the alphabet for the reactants and products of the rules.  This allows O(1) time access when updating the multiplicities of the proteins affected by the execution of a given rule, respectively.  We allow our algorithm to handle reactions of order no higher than three (trimolecular reactions).  Any higher order reaction can be broken down into subsequent smaller order reactions.  Hence, the Reactants and Products arrays are not larger than three.  This is important for the discussion of runtime at the end of the Section.

This concludes our discussion of Step I of the NWT algorithm.  The information of the Membrane System is completely contained within two arrays of structs: the Alphabet array and the Rules array.  Step II involves building the min-heap of the array.  For any two reactions, $R_1$ and $R_2$, in the heap, if $R_2$ is a child of $R_1$, we must have
\begin{equation}
WT_{R_1}\leq WT_{R_2}.
\label{minheapproperty}
\end{equation}

Steps I and II are each called only once during the simulation.  Now we discuss Step III of the NWT algorithm: Select Rule.  We want to select the reaction with the lowest Waiting Time.  Since we have organized our reactions in a min-heap, this step requires only $O(1)$ time to complete.  However, we must check to see if any other rules have the same Waiting Time -- i.\,e., reactions attempting to execute at the same exact instant.  These competing reactions could potentially be trying to use the same limited reactants.  If there are multiple rules slotted to occur at the same moment, then we must ensure there are enough molecules to satisfy all of the reactions.  If insufficient reactants exist, we will need to choose reactions nondeterministically until all available reactants have been exhausted.

We want to create a temporary array to store all the reactions with the minimum Waiting Time.  The elements of this array are pointers to nodes in the heap.  The first node in the array is the top of the heap.  We want to add all reactions to the array with the minimum Waiting Time.  To do so, we will recursively check children nodes until we stop finding tied reactions -- i.\,e., reactions attempting to occur simultaneously -- and build an array of pointers to these reactions.  



Once we build the array of ties (contains at least the top node in the min-heap), we move on to the next Step of the algorithm.  If there is only one element in the ties array, then there is no tie, and we can move on to Step V and apply only the one rule.  Otherwise, we must apply as many rules as possible in a nondeterministic manner.

To nondeterministically apply rules, we randomly generate numbers between 0 and the end of 
the ties array.  Using this randomly chosen index, we check if sufficient reactants exist 
to implement the reaction.  If there are sufficient reactants, we apply the 
reaction -- i.\,e., we increase the multiplicity of the product(s) by one and decrease the 
multiplicity of the reactant(s) by one.  If there are insufficient reactants, we skip the 
reaction, and no multiplicity changes occur for the reaction.  In either case, the reaction 
is removed from the ties array, and the process continues until the ties array is empty.  
This completes Step IV and Step V. Recall, in the case of only one reaction, we skip Step IV 
and apply just the one reaction in Step~V.  In either case, we are ready to move on to 
Step VI: Update Rules.

Step VI of the algorithm requires access to component (g) and (h) of the Reaction struct and 
component (d) of the Protein struct.  For each reaction applied in Step V, we must 
recalculate the Waiting Time of the applied reaction and the Waiting Time of every reaction
affected by the multiplicity changes.  We must discuss Step VI within the context of the 
heap maintenance.  Hence, we will continue the discussion of Step VI in 
Section \ref{heapmain}, which will also be a discussion of Step VIII, Heap Maintenance.  
As you will see in the next two Sections, Steps VI, VII, and VIII are all intertwined.  
But, the discussion on Step VII is left for Section \ref{memenhance}.

\subsection{Maintaining the min-heap}\label{heapmain}

As we stated earlier, we are building a min-heap from our Rules array with the bottom-up method.  However, the maintenance of the heap is accomplished in a nonstandard way.  Standard methods for heap maintenance  involve selecting the top node and removing it from the heap entirely.  Meanwhile, new nodes are added to the bottom and bubbled up as necessary.  There are a couple of reasons why we do not want to remove the nodes of applied rules from the heap and add them to the bottom.

For one, the number of reactions will not grow or shrink during a simulation run.  Hence, we do not need to remove or add reactions to the heap once it has been initialized.  Second, it is most often the case that, once a reaction is applied, its new Waiting Time is very close to the previous value.  Therefore, the top node will most likely be located near the top of the heap once the heap is resorted.  Hence, popping the top node and adding it to the bottom will often result in the node being bubbled back up to near the top of the tree (a waste of computer clock cycles, especially for a significantly large numbers of reactions).



For each affected rule (including the applied rule), we recalculate the WT (Step VI) and reposition the node in the heap (Step VIII).  This is accomplished one reaction at a time until all affected reactions are recalculated and repositioned as needed.

With the heap implementation, we were able to improve overall performance of the algorithm compared to our previous technique \cite{cheruku07}.   While the heap increased sorting, it also eliminated an extraneous FOR loop (running for every reaction in the tree) which was required to put Waiting Times in the context of simulation times.  The simulator described in \cite{cheruku07} has a runtime of $O(n^2\log{n})$.  In order to discuss the complexity of the NWT algorithm, there are several assumptions we make which are (usually) true for signaling cascades:

\begin{enumerate}
\item each reaction involves a maximum of five proteins;
\item the number of reactions having the same reactant is bounded (usually 3, at most 5);
\item due to the nature of chemical kinetics, the number of tied reactions is very small.
\end{enumerate}
From 1, 2 and 3 we say our algorithm has a runtime of $O(n\log{n})$ with respect to the number of reactions simulated.

This concludes our discussion of Steps VI and VIII.  We skipped Step VII, but we will now explain it in Section \ref{memenhance}.  Technically, Steps VI, VII and VIII all happen at the same time.  We see the interplay with Steps VI and VIII above, and Step VII merely factors into the recalculations of the Waiting Times.

\subsection{Memory enhancement}\label{memenhance}

In modeling many biochemical networks, situations will arise where one (or more) protein(s) 
($p_i\in \Sigma$) is a reactant for two or more reactions of different kinetic rates 
(fast vs. slow).  Consider the two simple diffusion laws for some molecule $A$ described by 
Table \ref{simptab}.

{\renewcommand{\arraystretch}{1.2}%
\begin{table}[h!t!p!]
\centering
\caption{An example system to illustrate memory enhancement.} \label{simptab}\vspace*{1mm}
\begin{tabular}{|l|c|c|}
\hline
Reaction & Rate Constant & Initial Molecules\\
\hline
$R_1$: $A\ra C$ & $k_1$ {\bf(slow)} & $A = 1$\\
\hline
$R_2$: $A\ra B$ & $k_2$ {\bf(fast)} & $B = 0$\\
\hline
& & $C = 0$\\
\hline
\end{tabular}
\end{table}}

This system is similar to the dynamics of the human immunodeficiencey type~1 (HIV-1) Tat protein.  If we assume $A$ is cytosolic Tat, $B$ is Tat in the nucleus, and $C$ is exocytised Tat, then we see that a molecule of Tat in the cytosol can be exocytised or translocated to the nucleus \cite{selliah01} according to the rules in Table~\ref{simptab}.  In the nucleus, Tat can upregulate HIV-1 proteins.  Outside the cell, Tat can affect neighboring cells.

We assume that $k_1<k_2$ and there is one molecule of $A$.  With these initial conditions, the NWT algorithm will send $A$ to $B$ for every single run.  Using the Gillespie Algorithm, there is a small chance $A$ will go to $C$.  Hence, after a significant number of simulations, the Gillespie Algorithm will yield a result of $A$ going to $C$. Modeling the system with differential equations will give the bizarre result of satisfying both rules, part of the molecule will satisfy $A$ goes to $B$ and part will satisfy $A$ goes to $C$.  For one molecule, this makes very little sense.

We use the concept of reaction memory to give deterministic results similar to ordinary 
differential equations (but maintaining molecular integrity).  With one molecule of $A$ 
in the system, we know \hbox{$WT_{R_1}>WT_{R_2}$}.  That is, $R_2$ is the next rule to be 
applied.  Once $R_2$ is applied, there are no more molecules of $A$ in the system.  
However, the next time a molecule of $A$ is in the system, we want $R_1$ to have a chance 
to apply.

When the Waiting Time of $R_1$ is changed to infinity (after the single molecule of $A$ changes into $B$), we store the percentage of time left to wait.  When a new molecule of $A$ becomes available to the system, the percentage is used in the recalculation of the Waiting Time of $R_1$.  The equation for a first order reaction is the following
\begin{equation}
WT_{R_3}=\mathit{Mem}\frac{1}{k_1*|A|}
\label{wtmem}
\end{equation}
\noi where $\mathit{Mem}$ is the percentage of time left to wait.

In a strictly deterministic sense, our algorithm is capable of generating equivalent results to an ordinary differential equations model.  But, with the nondeterminism of our algorithm, the memory enhancement can lead to different results.  For more information on the memory enhancement and results illustrating the improvement, see \cite{jack09}.

\section{Results}\label{results}

In this Section, we provide simulation results for two models, emphasizing the differences 
in results between the our technique, the Gillespie Algorithm, and solutions to systems of
ordinary differential equations.  The first model considered will be the Lotka-Volterra 
predator-prey model.  As a system of ordinary differential equations, we have
\begin{eqnarray}
\frac{dP_1}{dt}&=&P_1*(a-b*P_2)\nonumber\\      
\frac{dP_2}{dt}&=&-P_2*(c-d*P_1)        
\label{lotkaeqs}
\end{eqnarray}

\begin{figure}[ht]
\centering
(A)\includegraphics[width=4.6in]{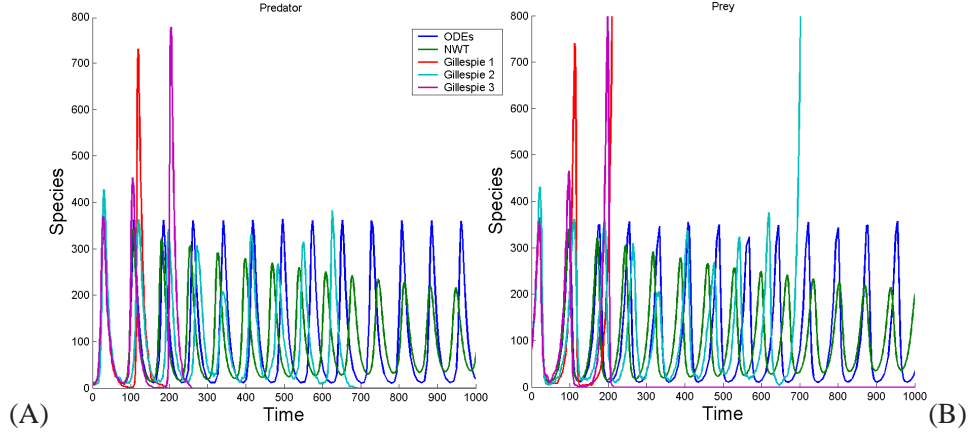}(B)
\caption{Results from predator-prey model.  (A) Predator species and (B) Prey species are both provided.}
\label{predprey}
\end{figure}

The model represents the interactions of two species: a predator population, $P_2$, and a 
prey population~$P_1$.  Prey species are born at a rate, $a$, and consumed at a rate,~$b$, 
while predator species are born at a rate, $d$, and die at a rate $c$.  
In Figure \ref{predprey}, we have two graphs, predator and prey, illustrating the results 
from simulations involving the NWT algorithm, the Gillespie Algorithm (three runs), and the 
system of ODEs.

Given a long enough time line, the Gillespie Algorithm will eventually result in no predator population, since every rule has a chance to occur at each time step.  For instance, runs 1, 2 and 3 show predator reaching zero at $t= $ $\sim195$, \hbox{$t= \sim250$}, and $t= $ $\sim690$ (see Figure \ref{predprey}(A)).  However, the NWT algorithm generates results comparable to the ordinary differential equations -- indefinite oscillations.  Although the oscillations for the NWT algorithm appear to be dampened, they do not approach zero but instead find an eventual steady pattern.  The reason the NWT algorithm does not produce the same exact results as the ordinary differential equations is molecular integrity.  Differential equations allow fractions of species to be considered, whereas the NWT algorithm only allows predator or prey species to increase/decrease by whole numbers. This difference is most likely responsible for the difference in amplitude of stable oscillations.

With the Lotka-Volterra model, we see the NWT algorithm has the capabilities to mimic the behavior of ordinary differential equations.  In the next model, we show how a small number of nondeterministic decisions can lead the NWT algorithm to generate results similar to the Gillespie Algorithm where ordinary differential equations fail to yield desireable results.

\begin{figure}[ht]
\centering
(A)\includegraphics[width=2.3in,height=2.3in]{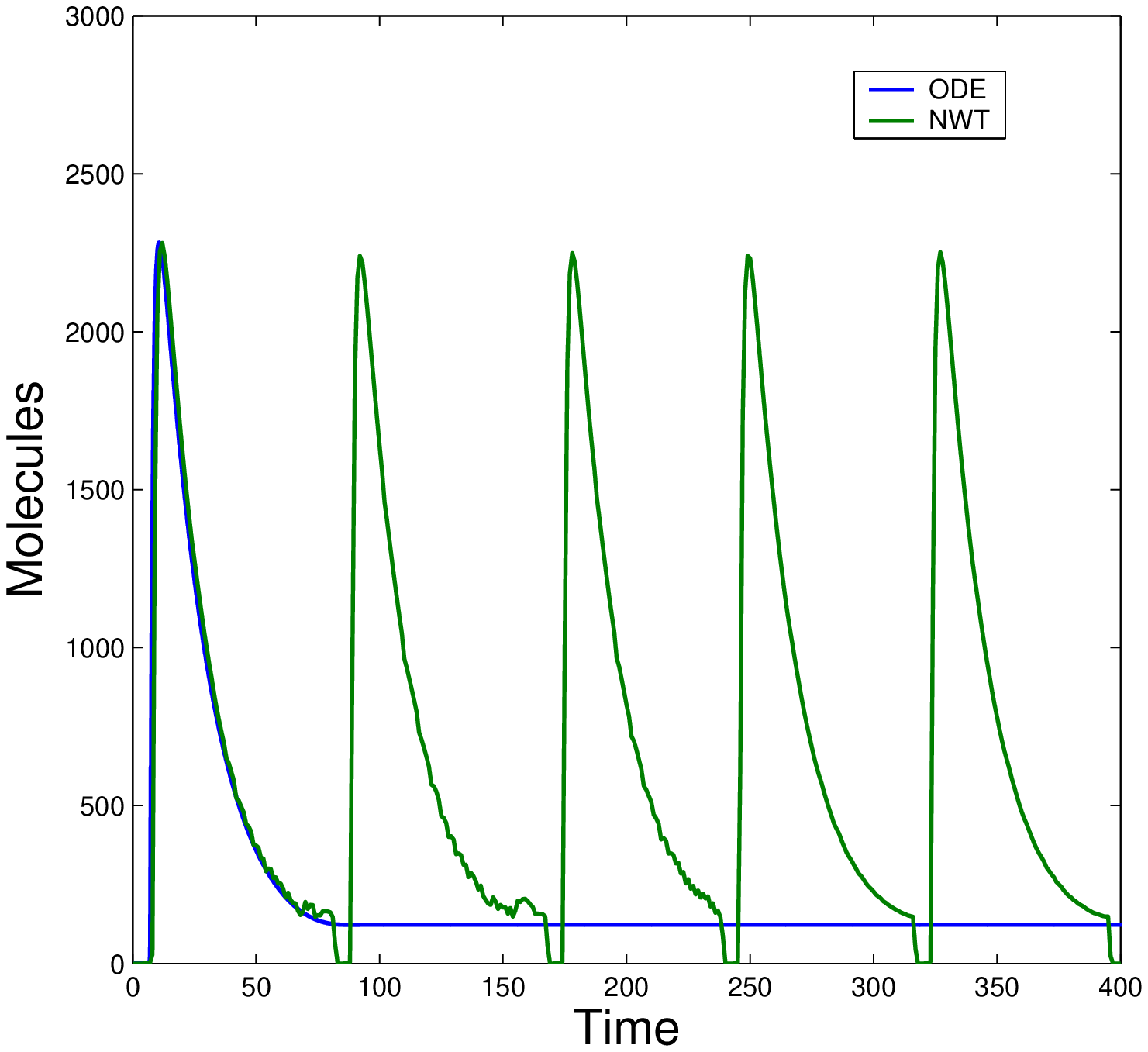}
\includegraphics[width=2.3in,height=2.3in]{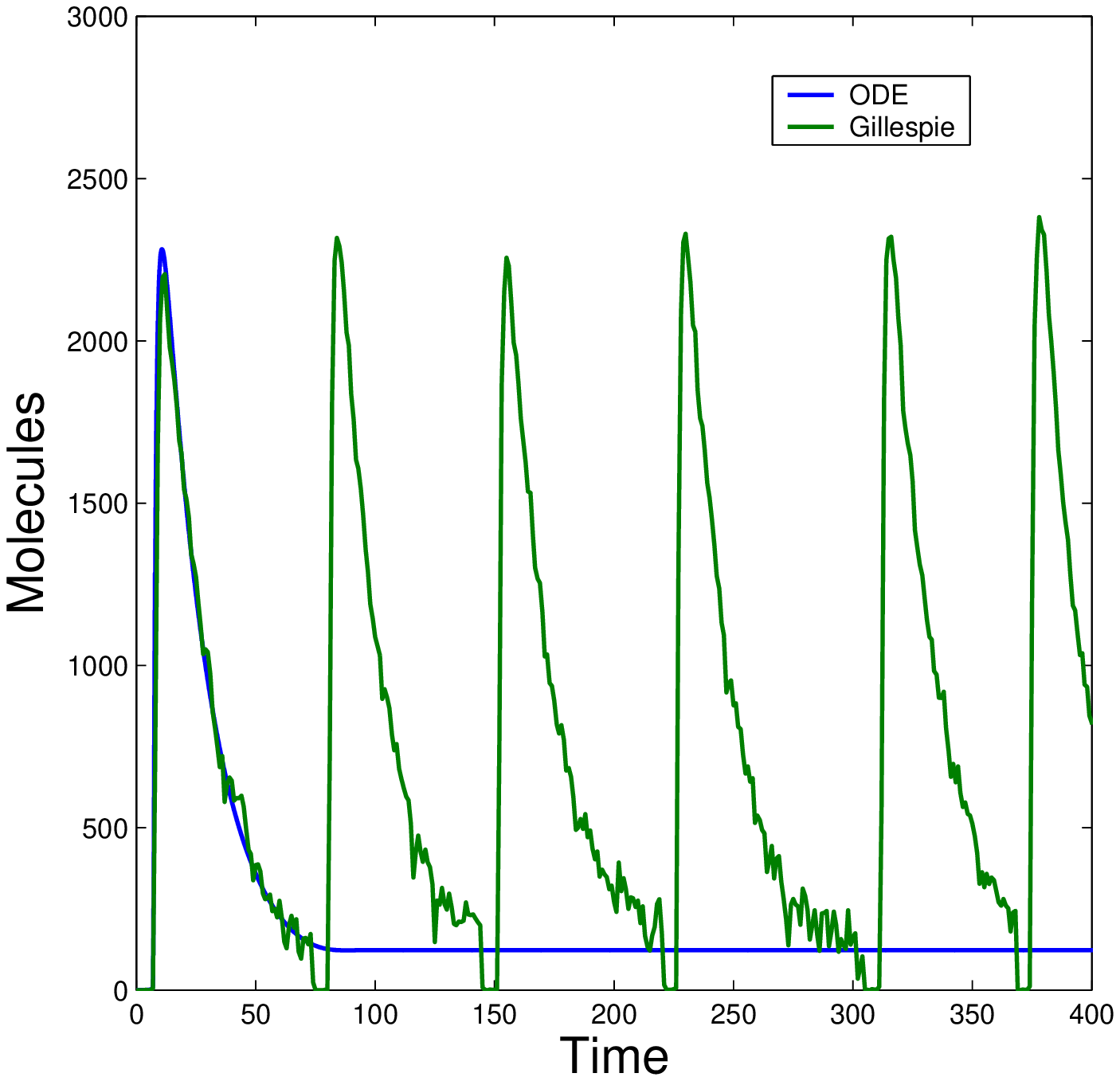}(B)
\caption{Results from circadian rhythm model.  (A) The NWT algorithm and (B) the Gillespie Algorithm exhibit a circadian rhythm whereas the ordinary differential equations reach a steady state after one initial peak.}
\label{circrhythm}
\end{figure}

We have chosen to model the circadian rhythm model described in \cite{vilar02}.  The model describes the behavior of an activator, $A$, and a repressor, $R$.  Transcription and translation rules are simulated as well as rules for the enhancement or inhibition of gene expression due to activator and repressor binding.  The system of ODEs is provided in Equation \ref{circEqs}.  The results of the simulations are shown in Figure \ref{circrhythm}.
\begin{eqnarray}
\frac{dD_A}{dt}&=&\theta_A*D'_A-\gamma_A*D_A*A\nonumber\\
\frac{dD_R}{dt}&=&\theta_R*D'_R-\gamma_R*D_R*A\nonumber\\
\frac{dD'_A}{dt}&=&\gamma_A*D'_R*A-\theta_A*D'_A\nonumber\\
\frac{dD'_R}{dt}&=&\gamma_R*D_R*A-\theta_R*D'_R\nonumber\\
\frac{dD_{M_A}}{dt}&=&\alpha'_A*D'_A+\alpha_A*D_A-\delta_{M_a}*M_A\nonumber\\
\frac{dA}{dt}&=&\beta_A*M_A+\theta_A*D'_A+\theta_R*D'_R-A*(\gamma_A*D_A+\gamma_R*D_R+\gamma_C*R+\delta_A)\nonumber\\
\frac{dM_R}{dt}&=&\alpha'_R*D'_R+\alpha_R*D_R-\delta_{M_R}*M_R\nonumber\\
\frac{dR}{dt}&=&\beta_R*M_R-\gamma_C*A*R+\delta_A*C-\delta_R*R\nonumber\\
\frac{dC}{dt}&=&\gamma_C*A*R-\delta_A*C
\label{circEqs}
\end{eqnarray}
\noi where $A$ and $R$ represent the number of activator and repressor proteins, $D'_A$ and $D_A$ represent the number of activator genes with or without binding to $A$, $D'_R$ and $D_R$ represent the number of repressor genes with or without binding to $R$, $M_A$ and $M_R$ represent mRNA molecules of $A$ and $R$, and $C$ represents the corresponding inactivated complex formed by $A$ and $R$.

In Figure \ref{circrhythm}, we see the similarities between the Gillespie Algorithm and the NWT technique.  Both algorithms are capable of producing sustained oscillations for the model, while a wholly deterministic approach (system of ODEs) results in one peak followed by a steady state.  The NWT algorithm is able to produce oscillations at a reduced dependence on nondeterministic decisions.  Although only one simulation run for each technique is shown in the graph, we provide the number of nondeterministic steps required for three runs per algorithm in Table~\ref{nondeter}.
\vspace{-1mm}

{\renewcommand{\arraystretch}{1.2}\begin{table}[h!t!p!]
\caption{Total Number of Nondeterministic Decisions for Time $= 2000$.\textcolor{white}{p}} \label{nondeter}\vspace{1mm}
\centering
\begin{tabular}{|l|c|c|}
\hline
 & Gillespie & NWT\\
\hline
Run 1 & 3071774 & 5093\\
\hline
Run 2 & 3029754 & 5185\\
\hline
Run 3 & 3103434 & 5435\\
\hline
\end{tabular}
\end{table}}

The Gillespie Algorithm makes a stochastic decisions for every single time step.  The NWT algorithm only makes a stochastic decision when reaction competition occurs over limited reactants -- two or more rules are trying to use a reactant which does not have enough molecules to satisfy those rules.  For all three runs of the NWT algorithm, less than $0.15\%$ of the total number of applied rules were nondeterminstically chosen, whereas $100\%$ of the applied rules for Gillespie were nondeterministically chosen.  Yet, the NWT algorithm is able to exhibit sufficient biochemical noise to induce oscillations similar to the Gillespie Algorithm.

\section{Conclusions and future work}\label{conclusions}

We have published several papers on our modeling technique.  In \cite{jack07,jack08}, we explore a model of the Fas-mediated apoptotic signaling cascades and compare the results with the modeling efforts of \cite{hua05}.  This work was extended in \cite{jack08hiv}, where we investigate the effects of HIV-1 proteins on the Fas-mediated signaling cascade within infected cells.  The study focuses on T cell latency, which is considered the last barrier to eradicate viral infection.  Also, in \cite{jack09} we discuss the memory enhancement and provide examples on its effectiveness.

In the future, we would like to further explore HIV-1 infection and extend the model 
described in~\cite{jack08hiv}.  We are also interested in modeling the effects of 
extracellular HIV-1 proteins on bystander cell apoptosis.  Furthermore, we will utilize the biomedical laboratories of Louisiana Tech University to deepen the understanding of the molecular mechanisms underlying Fas-induced apoptosis.  For instance, the effects of Ca$^{2+}$ on Fas-induced apoptosis.

Finally, we will look for ways to improve the NWT algorithm.  To date, we have only used MPI to run multiple simulations simultaneously.  However, we wish to explore methods where we partition a single Membrane System across multiple nodes, in an effort to increase the chances for reaction competition -- i.\,e., nondeterminism -- while maintaining efficiency.  We will also explore additional possibilities for nondeterminism (or stochasticity) to increase the chances for alternative biochemical evolutionary paths.

\bibliographystyle{eptcs}
\bibliography{paun}

\end{document}